\documentclass[twocolumn,preprintnumbers,secnumarabic,amsmath,amssymb,groupedaddress,nofootinbib]{revtex4}

\usepackage{dcolumn}% Align table columns on decimal point
\usepackage{bm}% bold math
\usepackage{mathtools}

\def\beq{\begin{equation}}
\def\eeq{\end{equation}}
\def\be{\begin{equation}}
\def\ee{\end{equation}}
\def\bea{\begin{eqnarray}}
\def\eea{\end{eqnarray}}

\usepackage{mathrsfs}
\usepackage{amsmath, amsthm, amssymb}
\usepackage{color}
\usepackage{epstopdf}
\usepackage[pdftex]{graphicx}
\usepackage{verbatim}
\usepackage{hyperref}
\usepackage{enumerate}
\usepackage{float}
\usepackage[caption = false]{subfig}
\usepackage[normalem]{ulem}
\usepackage{xfrac}

\begin{document}

\title{Superradiance Exclusions in the Landscape of Type IIB String Theory}

\author{Viraf M. Mehta$^{a}$}
\email{viraf.mehta@uni-goettingen.de}
\author{Mehmet Demirtas$^{b}$}
\author{Cody Long$^{b,c}$}
\author{David J. E. Marsh$^{d}$}
\author{Liam McAllister$^{b}$}
\author{Matthew J. Stott$^{d}$}

\vspace{1cm}
\affiliation{${}^a$ Institut f\"{u}r Astrophysik, Georg-August Universit\"{a}t, Friedrich-Hund-Platz 1, D-37077 G\"{o}ttingen, Germany}
\affiliation{${}^b$ Department of Physics, Cornell University, Ithaca, NY 14853, USA}
\affiliation{${}^c$ Department of Physics and CMSA, Harvard University, Cambridge, MA 02138, USA}
\affiliation{${}^d$ Department of Physics, Kings College London, Strand, London, WC2R, 2LS, United Kingdom}

\begin{abstract}

We obtain constraints from black hole superradiance in an ensemble of compactifications of type IIB string theory.
The constraints require knowing only the axion masses and self-interactions, and are insensitive to the cosmological model.
We study more than $2\cdot10^5$ Calabi-Yau manifolds with Hodge numbers $1\leq h^{1,1}\leq 491$ and compute the axion spectrum at two reference points in moduli space for each geometry.  Our computation of the classical theory is explicit, while for the instanton-generated axion potential we use a conservative model.  The measured properties of astrophysical black holes exclude parts of our dataset.  At the point in moduli space corresponding to the tip of the stretched K\"{a}hler cone, we exclude $\approx 50\%$ of manifolds in our sample at 95\% C.L., while further inside the K\"{a}hler cone, at an extremal point for realising the Standard Model, we exclude a maximum of $\approx 7\%$ of manifolds at $h^{1,1}=11$, falling to nearly zero by $h^{1,1}=100$.\\

\noindent KCL-PH-TH/2020-77
\end{abstract}

\maketitle

It has long been a dream of physicists working on quantum gravity and string theory to
test their theories against observational and experimental data. In this regard, string theory has proven rather stubborn.
Instead of a single low energy effective theory in four-dimensional spacetime, string theory provides a \emph{landscape} of possible theories, including compactifications of superstring theories on six-manifolds such as Calabi-Yau threefolds (CY$_3$'s)~\cite{1985NuPhB.258...46C} and discrete quotients thereof.
The effective theory depends on the topology of the internal space, for which there are astronomically many possibilities.
Testing string theory is therefore challenging for several reasons: the key phenomena are of gravitational strength, there is a vast set of theories to explore, and most predictions rest on model-dependent constructions of cosmology and of the visible sector.

In the present work, we develop a statistical test of part of the landscape based on the spectrum of axions, which can be related directly to the properties of the CY$_3$'s~\cite{1984PhLB..149..351W,2006JHEP...06..051S,2006JHEP...05..078C,axiverse,Marsh:2015xka,Demirtas:2018akl}.
We use the observed properties of astrophysical black holes to put limits on axions whose masses and self-interactions are in a range allowing for superradiant instabilities.  Such limits do not depend on the cosmological history, or on the detailed realization of the visible sector: they depend only on the Lagrangian of the dark sector, specifically that of axions, which we will compute in an ensemble of CY$_3$ compactifications.

{\textbf{Axions from the Kreuzer-Skarke Database:}} CY$_3$ hypersurfaces can be constructed from suitable triangulations of four-dimensional reflexive polytopes. A complete database of all such polytopes, numbering 473,800,776, was constructed by Kreuzer and Skarke~\cite{Kreuzer:2000xy}, and has been the subject of numerous studies~\cite{Altman:2014bfa,He:2015fif,Braun:2017juz,Altman:2017vzk,Cicoli:2018tcq,Huang:2018esr}.

Type IIB string theory contains a four-form field, $C_4$, in ten dimensions. Dimensional reduction of this four-form yields a number of axion-like fields (in the sense that they are pseudo-scalar phases)~\cite{Marsh:2015xka}, $\theta^i$:
\be
\theta^i := \int_{D^i} C_4\, ,
\ee
where $D^i$ is a closed four-dimensional submanifold. The size of a basis of such submanifolds is given by
the Hodge number $h^{1,1}$, so the index $i$ labelling axions takes on values from $1$ to $h^{1,1}$.
In the Kreuzer-Skarke list, one finds $1\leq h^{1,1}\leq 491$.
This class of solutions of string theory thus predicts an \emph{axiverse}~\cite{1984PhLB..149..351W,2006JHEP...06..051S,2006JHEP...05..078C,axiverse,Marsh:2015xka,Demirtas:2018akl}: a low energy theory containing a possibly large number of axions. The axions are one part of a complexified K\"{a}hler modulus field $T^i=\tau^i+i\theta^i$, with $\tau^i$ a K\"{a}hler modulus field.
%%%%%%%
\begin{figure*}
  \includegraphics[width=1.9\columnwidth]{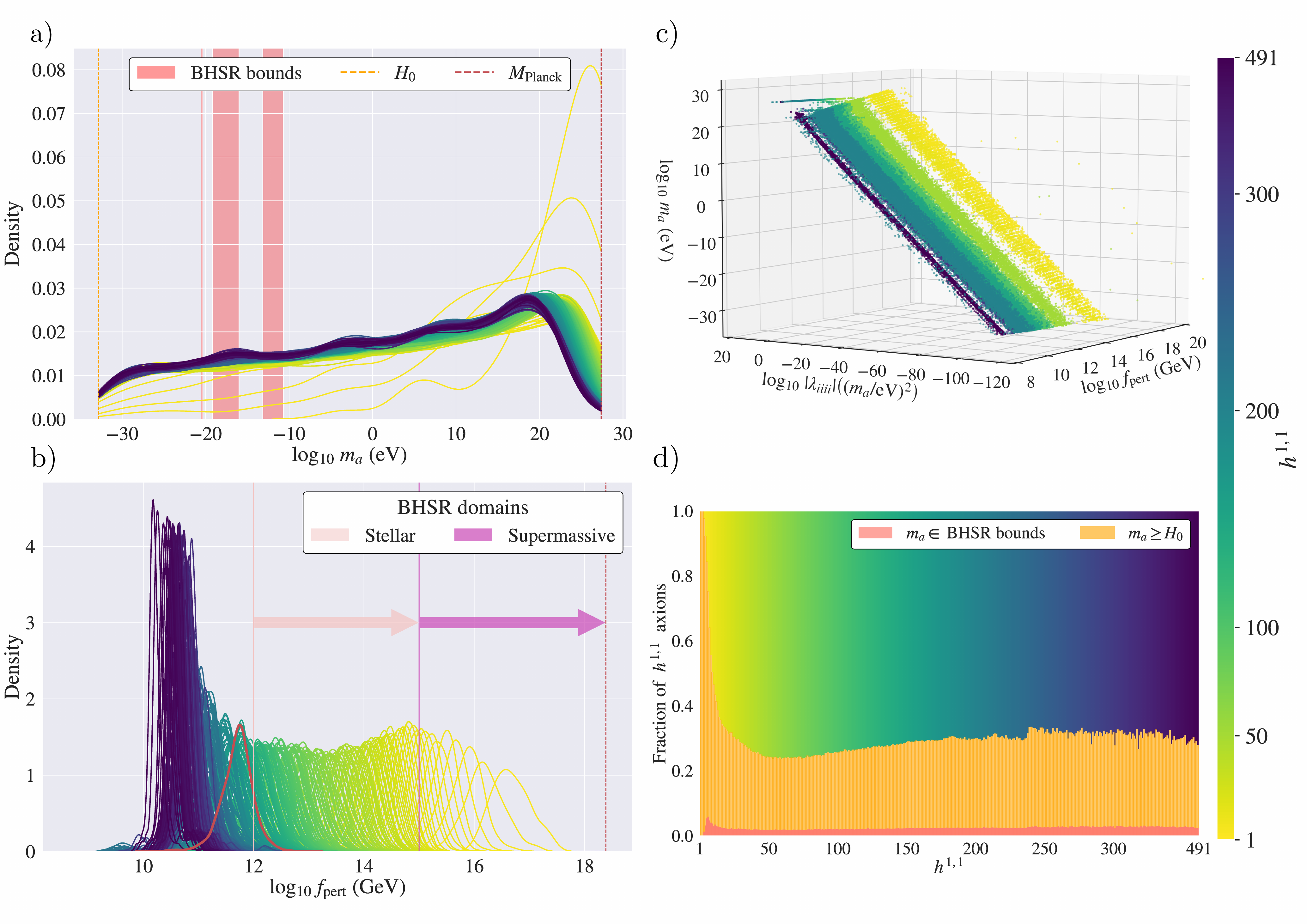}
\caption{\textbf{Summary Statistics:} We consider a number of manifolds at each value of the Hodge number, $h^{1,1}$, as described in the text. Summary statistics are binned by $h^{1,1}$ as indicated by the colour bar. a) Kernel Density Estimate (KDE) of the probability density of axion masses above the Hubble scale $H_0$, with the approximate BHSR region shaded.  b) KDE of axion decay constants. Indicated in red is the critical value $h^{1,1}=157$ (see text).  c) 3d scatter plot showing the derived physical data from a subset of $h^{1,1}$ values.
d) Bar plot showing the fraction of axions with $m\geq H_0$, and the fraction of axions in the BHSR region.}
\label{fig:stats_summary}
\vspace{-0.2in}
\end{figure*}
%%%%%%%%%%%

The axion fields have the Lagrangian:
\be
\mathcal{L} = -\frac{M_{\mathrm{pl}}^2}{8\pi^2}K_{ij}g^{\mu\nu}\partial_\mu\theta^i\partial_\nu\theta^j - \sum_{a=1}^\infty\,\Lambda_a^4\Bigl\{1-\cos\bigl(\mathcal{Q}^a{}_{i}\theta^i+\delta^a\bigr)\Bigr\}\, ,
\ee
where $g^{\mu\nu}$ is the inverse of the spacetime metric, $K_{ij}$ is the K\"{a}hler metric, and the second term is the instanton potential. The instanton potential contains energy scales $\Lambda_a$, charges $\mathcal{Q}^{a}{}_i$, and phases $\delta^a$.

Given a triangulation of a reflexive polytope from the Kreuzer-Skarke database, one can directly compute the K\"{a}hler metric as a function of the $\tau^i$.  For the instanton potential, we use a well-motivated model, namely that a generating set of holomorphic cycles --- specifically, prime toric divisors --- support the leading instantons \cite{Demirtas:2018akl}.  Then the scales $\Lambda_a$ are computable in terms of the topological data of the  CY$_3$ and the vevs of the moduli $\tau^i$.  A strong deviation from this model, involving for example dominant contributions from instantons on non-holomorphic cycles, would be a striking finding in its own right \cite{Demirtas:2019lfi}.  In this work we set the phases $\delta^a \to 0$, which is well-justified when the number of significant instantons is $\le h^{1,1}$.  We treat the general case in \cite{Mehta:2021pwf}.

The moduli fields, $\tau^i$, in general need to be stabilised, i.e.~given a potential and fixed to certain values. Schemes for this procedure are known for special cases~\cite{Kachru:2003aw,Balasubramanian:2005zx}, but the problem of moduli stabilisation is not solved in generality. In \cite{Demirtas:2018akl} and in the following, we simply examine the axion theory at specific points in the moduli space of the $\tau^i$.  The resulting theories may have light scalar fields $\tau^i$ and so are not necessarily realistic, but this does not preclude us from computing superradiance constraints from the axion sector at these points.

We place the moduli at specific locations in the \emph{stretched K\"{a}hler cone} (SKC), which is the region of moduli space within which the curvature expansion of string theory is well-controlled.  (As in \cite{Long:2016jvd}, these restrictions would need to be modified if the string coupling were extremely small.)  We consider two points in the SKC.  The first is the \emph{tip} of the SKC, i.e.~the point closest to the origin.
The second is an interior point defined by rescaling the K\"ahler parameters until the volume $\tau_{\rm{min}}$ of the smallest prime toric divisor  is $\tau_{\rm{min}} = 25 \approx 1/\alpha_{\rm{GUT}}$~\cite{Buras:1977yy,Dimopoulos:1981yj,pdg}, such that D7-branes wrapping this divisor could support a visible sector with a realistic grand unified gauge coupling $\alpha_{\rm GUT}$.  In more general constructions of the Standard Model the correct couplings may occur elsewhere in the SKC, but $\tau_{\rm{min}} \approx 25$ still provides a reasonable estimate of the point beyond which further dilation of the CY$_3$ would make the visible sector too weakly coupled.

{\textbf{Axion Spectra:}} We construct the axion potential for these two points in the SKC for $2 \cdot 10^5$ CY$_3$ hypersurfaces.
We include all favorable CY$_3$ hypersurfaces with $h^{1,1}\leq 5$, and for a random sample of 1000 hypersurfaces for every $6\leq h^{1,1}\leq 176$. The number of polytopes in the Kreuzer-Skarke list at each $h^{1,1}$ between 176 and 491 drops below 1000, and is sometimes zero, but we include at least 100 hypersurfaces for each $h^{1,1}$ in the list.

Using the optimisation suite \textsc{pygmo2}~\cite{Biscani2020}, we conducted minimisation using \textit{differential evolution} and searched for critical points and minima in the axion potentials for the studied triangulations.
We found that the axion statistics are remarkably robust across different local minima and critical points of the potential. Furthermore, all potentials with $\vec{\delta}=\vec{0}$ possess a critical point at $\vec{\theta}=\vec{0}$. Thus in the following we consider statistics at $\vec{\theta}=\vec{0}$ (for the general case see~\cite{Mehta:2021pwf}).

We construct the distributions of three quantities that describe the axion physics. First we compute the eigenvalues of $K_{ij}$ and evaluate $f_K :=M_{\mathrm{pl}}\sqrt{{\rm eigs}(K)}$.
The canonically normalised field $\tilde{\phi}^i$ is related to $\theta^i$ by $\theta^i=F^{i}{}_jU^{j}{}_{k}\tilde{\phi}^k$, where $F^i{}_{j}={\rm diag}[1/f_K]$ and $U^i{}_{j}$ is the unitary matrix that diagonalises $K_{ij}$. Next we compute the Hessian matrix $\mathcal{H}_{ij}$ and the tensor $\lambda_{ijkl}$ of fourth derivatives of the potential with respect to $\theta$.

Multiplying the Hessian by the transformation matrices to the canonical basis gives the mass matrix: $M_{ab}= \mathcal{H}_{ij}U^i{}_{k}U^j{}_{l}F^k{}_{a}F^l{}_{b}$. The eigenvalues of $M_{ij}$ give the axion masses-squared, $\{m_i^2\}$, and $M_{ij}$ is diagonalised by the unitary matrix $V^i{}_{j}$ defining the mass eigenbasis, $\phi^i$, from $\tilde{\phi}^i = V^i{}_{j}\phi^j$. Rotating to the canonically normalised mass eigenbasis leads to the interaction term in the Lagrangian $V_{\rm int} = \frac{1}{4!}\lambda^\phi_{ijkl}\phi^i\phi^j\phi^k\phi^l$, where $\lambda^\phi_{ijkl}$ is the axion self-interaction tensor in the canonical basis. We define the \emph{perturbative decay constant}, $f_{{\rm pert},i}$,
using the diagonal elements of $\lambda$ in the mass eigenbasis: $f_{{\rm pert},i}^2 :=m_i^2/\lambda_{iiii}$.

A statistical summary of our results for the axion masses $m_i$ and perturbative decay constants $f_{\rm pert}$ is shown in Fig.~\ref{fig:stats_summary} for moduli at the tip of the SKC. A full discussion of our results is presented in ~\cite{Mehta:2021pwf}. Summary statistics are collated for fixed $h^{1,1}$ only for presentation: for constraints we study each CY$_3$ separately.

As $h^{1,1}$ increases, the distribution of masses approaches an almost universal shape that is close to log-flat in the tails, with a bump near some characteristic large mass scale that becomes smaller as $h^{1,1}$ increases. The universal nature
of the tails can be seen, for example, from observing that as $h^{1,1}$ increases the fraction of massless axions, and of axions in any fixed mass window
below the bump, approach almost constant values. The decay constant $f_{\rm pert}$ follows an
approximately log-normal distribution, with the mean decreasing as $h^{1,1}$ increases. The log-normal distribution of $f_{\rm pert}$ can be understood from the product distribution of $m_i$ and $\lambda_{iiii}$ (which has the same shape as $m_i$) along with the strong, but not exact, correlation between $\lambda_{iiii}$ and $m_i$ visible in Fig.~\ref{fig:stats_summary}(c).

The trend of decreasing peak values of $f_{\rm pert}$ and $m_a$ with increasing $h^{1,1}$ can be understood geometrically from the increasing cycle sizes required by the greater topological complexity~\cite{Demirtas:2018akl}, and the inverse relationship between energy scale and volume characteristic of theories with extra dimensions.
The distribution of $f_{\rm pert}$ for an individual CY$_3$ displays mild scatter around the average for $h^{1,1}$.
Inside the SKC, the results are qualitatively the same, but with $f_{\rm pert}$ shifted down by approximately two orders of magnitude, and with the axion mass peak shifted down to near $10^{-10}\text{ eV}$, roughly consistent with $f_{\rm pert}^{-1} \sim -\mathrm{ln}\,m_a \sim \tau$.

%%%%%%%%%%%%
\begin{figure}
\includegraphics[width=0.99\columnwidth]{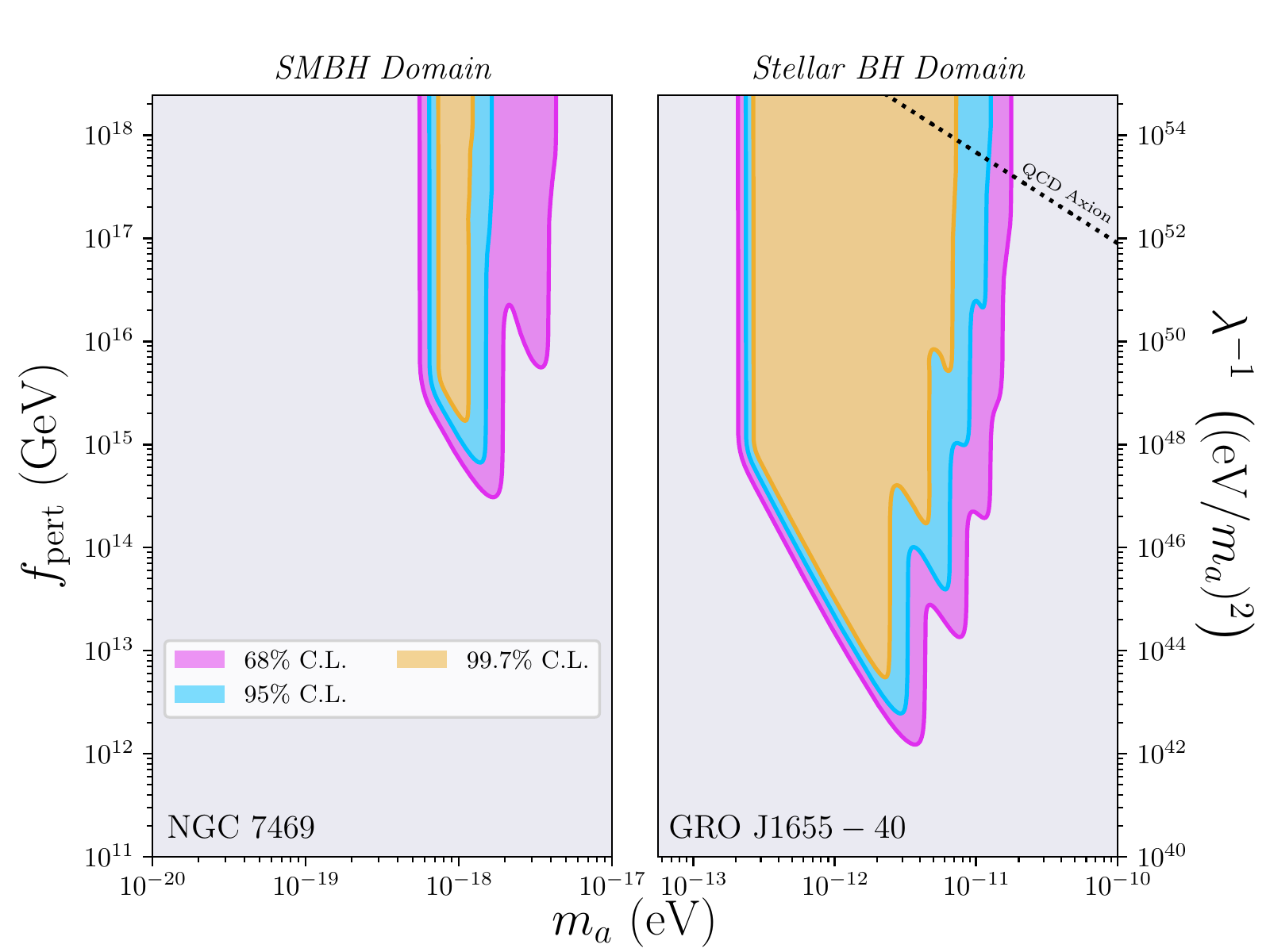}
\caption{Example exclusion functions for two typical BHs. The left panel shows a supermassive BH, and the right panel a stellar mass BH. BHSR operates over a resonant region in $m$, and is shut off by the Bosenova process at large values of the quartic coupling, parameterised by $f_{\rm pert}$.}
\vspace{-0.2in}
\label{fig:superradiance_exclusion}
\end{figure}
%%%%%%%%%%%%%

{\textbf{{Black Hole Superradiance (BHSR):}} Given the axion masses and quartic couplings, one can compute the effect of the axions on astrophysical black holes (BHs). A Kerr BH has an ergoregion in which timelike trajectories must co-rotate, leading to a spacelike time-translation Killing vector field external to the event horizon. The geometric nature of this region leads to growth of bosonic vacuum fluctuations~\cite{Pani:2012vp,Brito:2013wya,Brito:2020lup,Baryakhtar:2017ngi,Cardoso:2018tly,Siemonsen:2019ebd}. Quasi-bound-state modes, with a frequency $\omega$ satisfying the condition $\omega <\mu\Omega_{\rm H}$, with $\Omega_{\rm H}$ the angular velocity of the BH horizon and $\mu$ the angular momentum about the BH spin axis, return an associated negative Killing energy flux at the horizon. Energy conservation dictates that the external field source observed at spatial infinity must grow, at the cost of a reduction of the BH's angular momentum.

This process leads to
a superradiant instability, which is strongest when the axion Compton wavelength is approximately equal to the radius of the ergoregion. The evolutionary timescale, $\Gamma_{\rm SR}$, can be estimated via analytical approximations~\cite{Dolan:2007mj,1973ApJ...185..635T,PhysRevLett.29.1114,PhysRevD.22.2323,Baumann:2019eav,Arvanitaki:2010sy,Brito:2014wla,Brito:2015oca}.
Comparing these solutions to characteristic timescales for the evolution of a BH, and utilising measurements of known BH masses and spins, leads to exclusions on the axion mass, $m$.

Non-linear phenomena \cite{Arvanitaki:2010sy,Arvanitaki:2014wva,Baumann:2018vus,Zhang:2019eid,PhysRevD.99.064018,Baumann:2019ztm,Berti:2019wnn,Kavic:2019cgk,Cardoso:2020hca} may inhibit the exponential amplification of the dominant state, quenching the instability.
For example, when the self-interactions between the bosons are attractive, the cloud undergoes a rapid collapse known as a \emph{Bosenova}, shutting down the superradiant instability at a critical occupation number proportional to the strength of the quartic coupling \cite{Arvanitaki:2010sy,Arvanitaki:2014wva,Yoshino:2012kn,Yoshino:2015nsa,Mocanu:2012fd}.
More generally, it is possible to estimate a critical cloud size when interactions of any form outcompete the superradiant growth.
The tensor $\lambda_{ijkl}$ contains off-diagonal components that allow axions in the superradiant cloud to decay and annihilate to lighter axions, which could compete with superradiance.  We find that the off-diagonal components are smaller than the diagonal ones by many orders of magnitude, and so these processes can be neglected.  This leads to a two-dimensional exclusion function in the domain of the axion mass and self-interaction, with the latter parameterised by $f_{\rm pert}$.

We adopt the detailed model for superradiance, and the BH data compilations, presented in ~\cite{Stott:2018opm,Stott:2020gjj,Mehta:2021pwf}.
We pre-compute the exclusion probability for each BH in the plane $(m,f_{\rm pert})$: an example is shown in Fig.~\ref{fig:superradiance_exclusion} for two typical BHs. For any manifold, once the set of $(m,f_{\rm pert})$ is determined, then the model is excluded if even \emph{one} axion falls into the exclusion region.

We repeated our analysis excluding both supermassive BHs and those with spins estimated from gravitational waveforms \cite{Mehta:2021pwf}, and found only a small change in our constraints.  The primary reasons are that for supermassive black holes the self-interactions required to quench superradiance are weaker, and for BHs with gravitational wave spins the axion mass ranges covered by these BHs already largely overlap others in the sample.  Thus, our conclusions are dominated by the stellar-mass BHs whose spins are measured using X-rays in their accretion disks.

{\textbf{{Constraints on the Landscape:}} Fig.~\ref{fig:exclusions} shows our main result: the fraction of excluded CY$_3$'s at each value of $h^{1,1}$ for moduli at the tip of the SKC, and for moduli inside the SKC. We show the constraints both with and without the effect of self-interactions, illustrating the importance of the latter.  The self-interactions lead to a weakening of constraints at higher $h^{1,1}$, where the trend for lower $f_{\rm pert}$ leads to stronger self-interactions, and thus a higher probability that interactions will quench superradiance.

%%%%%%%%%%
\begin{figure*}
\includegraphics[width=1.9\columnwidth]{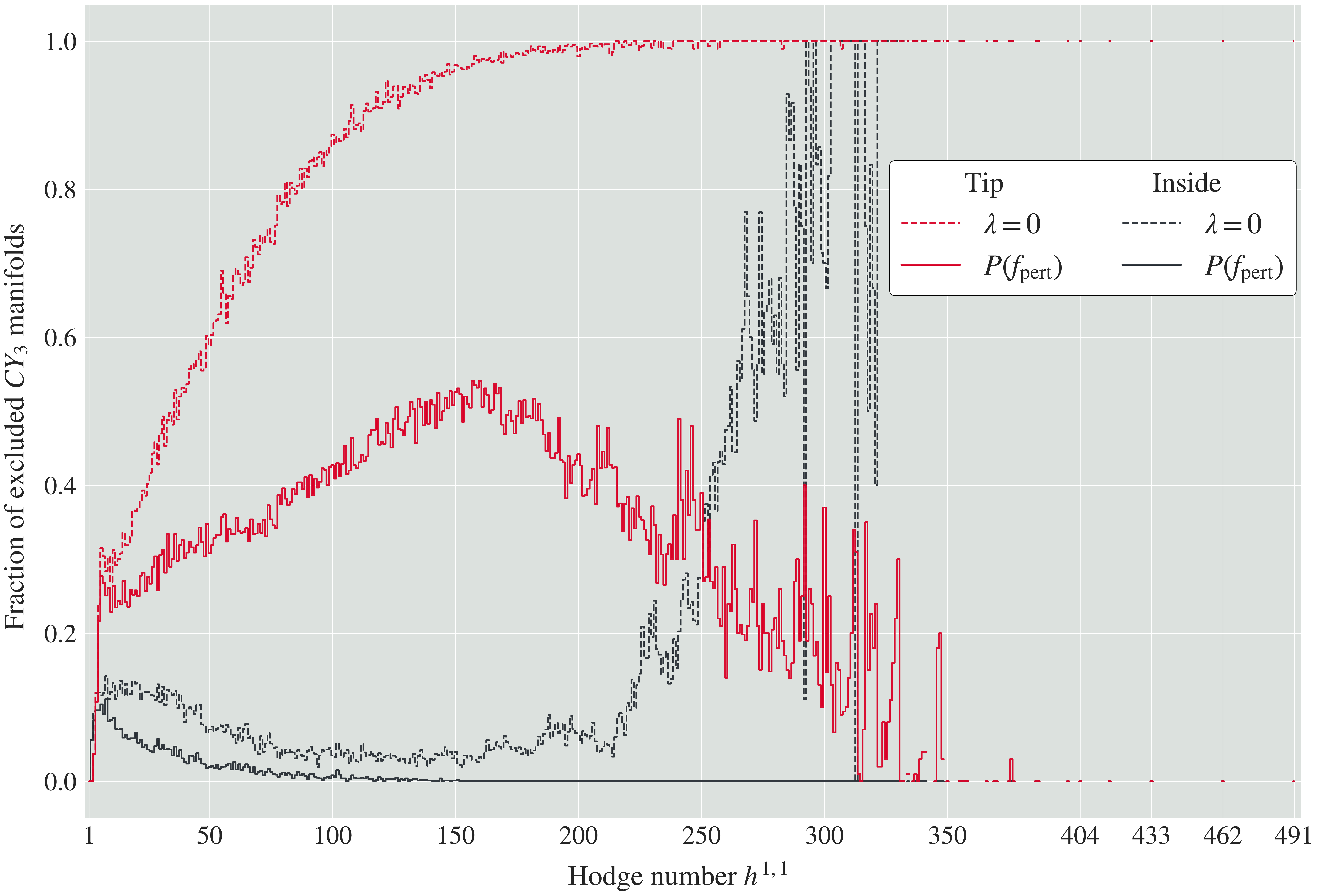}
\caption{\textbf{Fraction of Calabi-Yau Hypersurfaces Excluded by Black Hole Superradiance}. Dashed lines omit self-interactions, while solid lines include self-interactions and the resulting possibility of a Bosenova.  Red lines are at the tip of the SKC, and black lines are evaluated inside the SKC.}
\vspace{-0.2in}
\label{fig:exclusions}
\end{figure*}
%%%%%%%%%%%

At the tip of the SKC, the excluded fraction of CY$_3$'s rises rapidly with $h^{1,1}$ as the mass distribution spreads to encompass the BHSR region and the number of axions grows, reaching $\approx 50\%$ at $h^{1,1}\approx 157$.  For $h^{1,1} > 157$, $f_{\mathrm{pert}}$ decreases, so that self-interactions eventually shut off BHSR and cause the constrained fraction to fall, reaching zero for $h^{1,1} \gtrsim 350$.

Inside the SKC, the axion masses are already much lighter due to the increased cycle volumes, with the bulk of the distribution lying below the BHSR region. Similarly, the decay constants are two orders of magnitude smaller than at the tip, such that self-interactions shut off the BHSR bounds at a lower value of $h^{1,1}$.  The exclusion fraction of CY$_3$'s reaches a maximum of only $\approx 7\%$ at $h^{1,1}=11$, and declines to almost zero by $h^{1,1}=100$.  Thus, at this point in moduli space, CY$_3$'s with large $h^{1,1}$ are essentially unconstrained by BHSR.

{\textbf{{Validation of our Methodology:}} Constructing our ensemble of axion theories involved sampling points in the K\"ahler cones of CY$_3$'s, as well as
modeling the charges $\mathcal{Q}^a_{~i}$.  We tested the effects of changes in these steps: we sampled an enlarged K\"ahler cone, we used two different sampling algorithms, and we included contributions to $\mathcal{Q}^a_{~i}$ from instantons wrapping linear combinations of prime toric divisors, including non-holomorphic combinations.  We also included random $CP$ phases $\delta^a$.  As shown in detail in~\cite{Mehta:2021pwf}, none of these tests led to a meaningful difference in our exclusions.

{\textbf{Moving Beyond BHSR:}} Superradiance constraints rely only on gravitational interactions and vacuum fluctuations, and so provide a comparatively model-independent test of the string landscape~\cite{axiverse,Marsh:2015xka}.
However, for very large $h^{1,1}$ other constraints may become important.
In a sample of 10,000 geometries with $h^{1,1}=491$ \cite{Demirtas:2020dbm}, the mean $f_K$ was $\langle f_K\rangle\approx10^{10}\text{ GeV}$ at the tip of the SKC, which is consistent with the present analysis.  The axion-photon coupling is $g=c_{\rm mix}\alpha_{\rm em}/2\pi f_K$, where $\alpha_{\rm em}$ is the electromagnetic fine structure constant, and $c_{\rm mix}\sim \mathcal{O}(1)$ arises from mixing between dark and visible $U(1)$'s. Using $\langle f_K\rangle$ in this estimate leads to values consistent with those found in \cite{Halverson:2019cmy}. A visible sector coupling of this magnitude for a massless axion is close to current astrophysical constraints~\cite{Payez:2014xsa,Day:2018ckv,Reynolds:2019uqt}, which demand $g\lesssim 10^{-12}-10^{-13}\text{ GeV}^{-1}$. This suggests that further study of visible sector couplings at large $h^{1,1}$ could lead to significant constraints on the landscape.

Our results show that it is possible to make quantitative progress in constraining  the landscape.  The more challenging problem is to look for evidence in favour of string theory in the remaining --- and still vast --- parts of the landscape. The axion spectra we have computed may hold the answers.

{\textbf{{Acknowledgements:}} We thank M.~Kim, J.~Moritz, and J.~Stout for discussions.  VMM is, and DJEM was, supported by the Alexander von Humboldt Foundation and the German Federal Ministry of Education and Research. DJEM is supported by an Ernest Rutherford Fellowship from the UK STFC. The work of MD and LM is supported in part by NSF grant PHY-1719877. The work of CL is supported in part by the Alfred P. Sloan Foundation Grant No. G-2019-12504. The work of MJS is supported by funding from the UK Science and Technology Facilities Council (STFC).
This work made use of the Scientific Computing Cluster at the University of G\"ottingen \cite{gwdg} and the open source packages \textsc{cytools}~\cite{cytools}, \textsc{matplotlib}~\cite{matplotlib}, \textsc{mpmath}~\cite{mpmath}, \textsc{numpy}~\cite{numpy}, \textsc{pandas}~\cite{mckinney-proc-scipy-2010,reback2020pandas}, \textsc{scipy}~\cite{scipy}, and \textsc{seaborn}~\cite{waskom2020seaborn}.

\bibliography{KSAxiverseBHSR}

\end{document}